\documentclass[a4paper, amsfonts, amssymb, amsmath, reprint, showkeys, footinbib, ,twoside,superscriptaddress,floatfix,longbibliography]{revtex4-1}

\usepackage{graphicx} 
\usepackage{amsmath}
\usepackage{comment}
\usepackage{booktabs}
\usepackage{hyperref}
\usepackage{url}
\usepackage{adjustbox}
\usepackage{array}
\usepackage{xcolor}

\newcommand{\rrev}[1]{{\color{black} #1 }}
\newcommand{\rev}[1]{{\color{black} #1 }}

\newcommand{\figLabelCapt}[1]{\textbf{\MakeLowercase{{#1}}}}
\newcommand{\refSub}[2]{\hyperref[#2]{\ref{#2}\figLabelCapt{#1}}}

\begin{document}

\title{Cartesian atomic cluster expansion \\
for machine learning interatomic potentials}

\author{Bingqing Cheng}
\email{bingqingcheng@berkeley.edu}
\affiliation{Department of Chemistry, University of California, Berkeley, CA, USA}
\affiliation{The Institute of Science and Technology Austria, Am Campus 1, 3400 Klosterneuburg, Austria}

\date{\today}

\begin{abstract}
Machine learning interatomic potentials are revolutionizing large-scale, accurate atomistic modelling in material science and chemistry. Many potentials use atomic cluster expansion or equivariant message passing frameworks. Such frameworks typically use spherical harmonics as angular basis functions, \rrev{followed by Clebsch-Gordan contraction to maintain rotational symmetry. We propose a mathematically equivalent and simple alternative that performs all operations in the Cartesian coordinates}. This approach provides a complete set of polynormially independent features of atomic environments while maintaining interaction body orders. Additionally, we integrate low-dimensional embeddings of various chemical elements, \rrev{trainable radial channel coupling,} and inter-atomic message passing. The resulting potential, named Cartesian Atomic Cluster Expansion (CACE), exhibits good accuracy, stability, and generalizability. We validate its performance in diverse systems, including bulk water, small molecules, and 25-element high-entropy alloys.
\end{abstract}

\maketitle


Machine learning interatomic potentials (MLIPs) can learn from quantum-mechanical calculations and predict the energy and forces of atomic configurations speedily, thus enabling more precise and comprehensive exploration of material and molecular properties at scale~\cite{keith2021combining,unke2021machine}.
Crucially, while atomic Cartesian coordinates encode all the essential information of a structure, 
they cannot be directly used in learning tasks,
due to the lack of symmetries like translation, rotation, inversion, and atom permutations.
Consequently, numerous methods have been developed to enforce these symmetries~\cite{musil2021physics,unke2021machine}.

In particular, atomic cluster expansion (ACE)~\cite{drautz2019atomic} 
represents atomic environments based on a body-order expansion, using a complete set of bases of
spherical harmonics and radial components.
ACE can be viewed as a general framework of many other representations of the atomic environment, such as the Atom Centered Symmetry Functions (ACSF)~\cite{behler2007generalized}, Smooth Overlap of Atomic
Positions (SOAP) descriptor~\cite{bartok2010gaussian}, Moment Tensor Potentials (MTPs)~\cite{shapeev2016moment}, 
and bispectrums~\cite{bartok2013representing}. 

Another important approach to learning atomic interactions uses message passing neural networks (MPNNs).
They represent structures as graphs with atoms as nodes,
and apply message passing operations to
learn atomic environment representations.
Earlier MPNN models, such as SchNet~\cite{schutt2017schnet}, PhysNet~\cite{unke2019physnet},
SphereNet~\cite{liu2022spherical} and GemNet~\cite{gasteiger2021gemnet},
use internal features that are invariant under rotations.
In models such as NewtonNet~\cite{haghighatlari2022newtonnet}, EGNN~\cite{satorras2021n}, and PaiNN~\cite{schutt2021equivariant}, vector features built using relative atomic positions are also used on top of invariant features.
In later equivariant MPNNs, such as Cormorant~\cite{anderson2019cormorant}, NequIP~\cite{batzner20223}, MACE~\cite{batatia2022mace} 
exploit internal features based on irreducible representations of the E(3) symmetry group~\cite{geiger2022e3nn,goodman2000representations}.
\rrev{E(3) equivariant MPNNs were able to achieve an unprecedented accuracy compared to the previous invariant architectures~\cite{batzner20223,batatia2022mace}.}

Under the hood, in both ACE and E(3) equivariant MPNNs, 
the key process is the following:
\rev{Equivariant features or messages of body order $\nu$ are encoded in
spherical harmonics with degrees $l$.
The invariant features are subsequently created by contracting the equivariant ones via the Clebsch-Gordan coefficients.}
These invariants are used to predict energies or other invariant physical quantities.
\rrev{
The use of spherical harmonics is motivated by that they forms a natural basis for the irreducible representations of the SO(3) group and thus is useful in operations involving rotational symmetries.
Since the original ACE work~\cite{drautz2019atomic}, various improvements have been made, e.g.,
efficient implementations of spherical harmonics that are evaluated in Cartesian coordinates are made~\cite{sloan2013efficient,drautz2020atomic,sphericart},
absorbing Clebsch-Gordan matrix in expansion coefficients during the evaluation stage~\cite{lysogorskiy2021performant}, 
recursive Clebsch-Gordan coupling~\cite{batatia2022mace} or using Gaunt coefficients~\cite{luo2024enabling}.
Moreover, the original formulation has a steep scaling as the number of elements increases,
and tensor-reduced representations have been developed to combat this~\cite{darby2023tensor}.
As the degree $l$ and the body order $\nu$ increases, 
many invariant features are linearly dependent or polynomially depedent, and schemes to eliminate these have been developed~\cite{dusson2022atomic,nigam2020recursive,goff2024permutation}.
}

\rrev{On the other hand, MTP~\cite{shapeev2016moment} can be considered as ACE formulated in the Cartesian space.
MTP uses the so-called moment tensors, and then contracts them to form rotationally invariant basis functions~\cite{shapeev2016moment}.
The MTP contraction yields a spanning set, but contains many linearly dependent basis functions~\cite{shapeev2016moment,dusson2022atomic}.
In a similar vain, proper orthogonal descriptors were introduced as compact orthogonal basis functions formulated in the Cartesian space~\cite{nguyen2023fast}.
}

Here, we propose an alternative method that performs the expansion of atomic density as well as \rrev{the contraction to get polynormially indepedent invariant features} in Cartesian coordinates directly,
circumventing spherical harmonics.
The Cartesian atomic cluster expansion (CACE) has the same mathematical foundation as the ACE framework, so it also adopts the nice properties of ACE: body order and completeness in the description of atomic environments~\cite{dusson2022atomic}.
The CACE invariant features are low-dimensional and can be computed independently and efficiently.
By further incorporating a low-dimensional embedding for chemical elements, \rrev{optimized radial channel coupling,} and message passing between neighboring atoms,
we develop the CACE potential. 
We benchmark CACE on diverse systems,
with an emphasis on stability and extrapolation.

\section{Results}

\subsection{The CACE architecture}

\begin{figure*}
  \centering
  \includegraphics[width=0.95\textwidth]{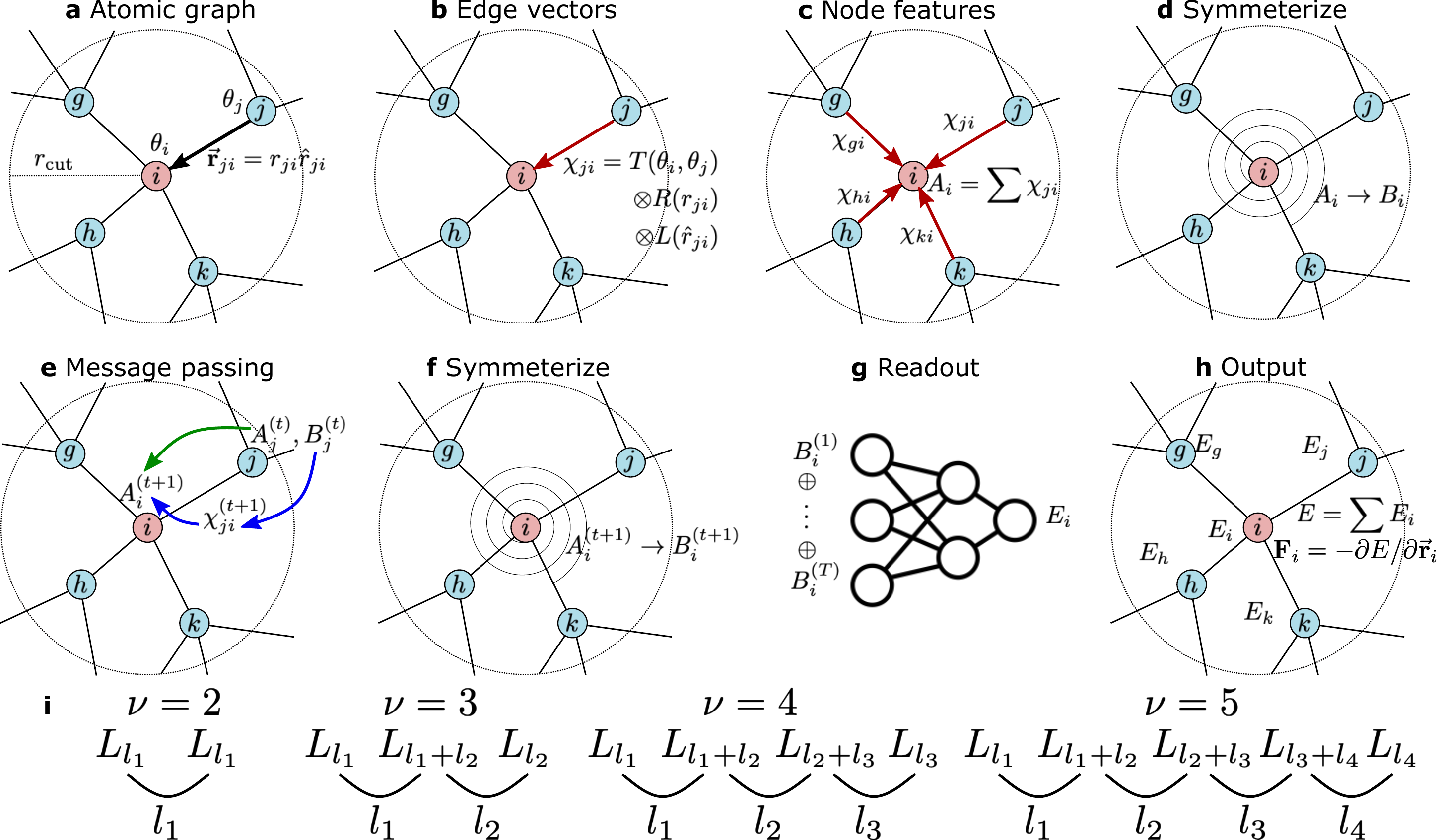}
    \caption{
    Schematic of the CACE potential. \textbf{a}-\textbf{h} show each step of the operation, and \textbf{i} illustrates the rule for making rotationally invariant features.
    }
    \label{fig:schematics}
\end{figure*}

\textbf{Atomic graph}
As illustrated in Fig.~\ref{fig:schematics}, each atomic structure is represented as a graph, 
with atoms as nodes, and directed edges that point from pairs of sender atoms to receiver atoms that are within a cutoff radius of $r_\mathrm{cut}$.
The state of an atom $i$ is denoted by the tuple $\sigma_i = (\textbf{r}_i, \boldsymbol{\theta}_i, \textbf{h}_i)$:
$\textbf{r}_i$ is its Cartesian position,
$\boldsymbol{\theta}_i$ encodes its chemical element $\alpha$, and $\textbf{h}_i$ is the learnable features.
Later we will use the superscript, e.g. $\sigma_i^{(t)}$, to denote the atomic state in the message passing layer $t$, but for now we omit the layer-dependence indices for simplicity.

For both the sender node and the receiver node of an edge, elements $\alpha_i$ are first assigned to vectors of lengths equal to the total number of elements $N_\mathrm{element}$ through a one-hot embedding,
and then the one-hot vectors are multiplied with a learnable weight
matrix of size $N_\mathrm{embedding} \times N_\mathrm{element}$ that outputs a learnable embedding $\boldsymbol{\theta}_i$ for each atom $i$ of length $N_\mathrm{embedding}$.
Typically, $N_\mathrm{embedding}$ is selected to be a small number between 1 and 4.
This continuous element embedding eliminates the unfavorable scaling of using 
discrete chemical element types when representing atomic environments, and also allows alchemical learning.

\textbf{Edge basis}
The edge basis, shown in Fig.~\ref{fig:schematics}b, is used to describe the
spatial arrangement of an atom $j$ around the atom $i$ 
($\vec{\mathbf{r}}_{ji} \equiv \mathbf{r}_{j} - \mathbf{r}_{i}$, $r_{ji} = | \vec{\mathbf{r}}_{ji} | $
and $\hat{\mathbf{r}}_{ji} = \vec{\mathbf{r}}_{ji} / r_{ji}$),
and is formed by the product of a radial basis $R$,
an angular basis $L$, and an edge-type basis $T$:
\begin{equation}
    \chi_{cn\mathbf{l}} (\sigma_i, \sigma_j) =  
    T_{c}(\sigma_i, \sigma_j)
    R_{n,cl}(r_{ji})
    L_{\mathbf{l}}(\hat{\mathbf{r}}_{ji}).
    \label{eq:edge-basis}
\end{equation}
Eqn.~\eqref{eq:edge-basis} is general and fundamental for many MLIPs, and
a systematic discussion of the design choices for each term can be found in Ref.~\cite{batatia2022design}.

In CACE, the type of edge $T$ depends on the states of the two nodes it connects.
In the initial layer,
the edge states are only determined by the chemical elements of the two atoms $i$ and $j$.
The edge is encoded using the flattened tensor product of the two node embedding vectors, $T=\boldsymbol{\theta}_i \otimes \boldsymbol{\theta}_j$,
resulting in an edge feature of length $c=N_\mathrm{embedding}^2$.
In a later message passing layer $t$, the edge features can depend on the node hidden features $\mathbf{h}_i^{(t)}$.
This edge encoding formed by the tensor product can be interpreted as the attention mechanism~\cite{vaswani2017attention} between the two atoms,
with the embedding vectors of the sender and the receiver nodes the query and the key.

The orientation-dependent angular basis is
\begin{equation}
    L_{\mathbf{l}}(\hat{\mathbf{r}}_{ji}) =
    (x_j-x_i)^{l_x} (y_j-y_i)^{l_y} (z_j-z_i)^{l_z},
    \label{eq:angular-basis}
\end{equation}
where $\mathbf{l} \equiv (l_x, l_y, l_z)$ are the angular momentum numbers satisfying $l_x + l_y + l_z = l$, with $l$ being the total angular momentum.
Such angular basis is just spherical harmonic functions in Cartesian basis.
Indeed, one can easily transform the function of a certain angular momentum number $l$ in Cartesian and spherical harmonic bases by a simple matrix multiplication~\cite{altmann1957symmetries}.
As examples of operations in the Cartesian angular basis,
one can add two vectors such as
\begin{equation}
    L_{\mathbf{l}_3}(\hat{\mathbf{r}}_{ji}+\hat{\mathbf{r}}_{ik}) =
    \sum_{\mathbf{l}_1+\mathbf{l}_2=\mathbf{l}_3}
\mathcal{P}(\mathbf{l}_1, \mathbf{l}_2)
    L_{\mathbf{l}_1}(\hat{\mathbf{r}}_{ji})
    L_{\mathbf{l}_2}(\hat{\mathbf{r}}_{ik})
    \label{eq:l-add}
\end{equation}
where the prefactor $\mathcal{P}(\mathbf{l}_1, \mathbf{l}_2)=\dfrac{l_{3x}!}{l_{1x}!l_{2x}!}
    \dfrac{l_{3y}!}{l_{1y}!l_{2y}!}
    \dfrac{l_{3z}!}{l_{1z}!l_{2z}!}$
    with $\mathbf{l}_1+\mathbf{l}_2=\mathbf{l}_3$.
One can obtain vector dot products such as
\begin{equation}
    (\hat{\mathbf{r}}_{ji} \cdot \hat{\mathbf{r}}_{ki})^l
    = \sum_{\mathbf{l}} \mathcal{C}(\mathbf{l}) 
    L_{\mathbf{l}}(\hat{\mathbf{r}}_{ji})
    L_{\mathbf{l}}(\hat{\mathbf{r}}_{ki}),
    \label{eq:l-dot}
\end{equation}
where
$\mathcal{C}(\mathbf{l}) = \dfrac{l!}{l_x ! l_y ! l_z !}$ is the combinatorial coefficient, and the
sum is over all $\mathbf{l}$ with $l_{x}+l_{y}+l_{z} =l$.
One can prove the relationships in Eqns.~\eqref{eq:l-add} and ~\eqref{eq:l-dot} above using the multinomial formula.

$R_{n,cl}(r_{ji})$ is a set of radial basis consisting of $n$ functions, coupled with the total angular momentum $l$ and the edge channel $c$.
\rrev{
CACE first uses a set of raw radial basis
$R_{\Tilde{n}}(r_{ji})$ consisting of $\Tilde{n}$ functions, which can be a set of trainable Bessel functions with different frequencies~\cite{schutt2021equivariant} multiplied by a smooth cutoff function. 
The raw radial basis is mixed using a linear transformation to form $R_{n,cl}$ for each $c$ and $l$, 
\begin{equation}
    R_{n,cl}(r_{ji}) = \sum_{\Tilde{n}} R_{\Tilde{n}}(r_{ji}) W_{\Tilde{n}n,cl}
    \label{eq:r-mix}
\end{equation}
where $W_{\Tilde{n}n,cl}$ is a $\Tilde{n}\times n$-sized matrix indexed by $c$ and $l$.
This means that, for each $l$ and $c$ combination, there is a different set of $n$ radial functions.
In practice, the mixing is performed on the atom-centered basis for efficiency, as described below.
}

\textbf{Atom-centered basis ($A$-basis)}
As shown in Fig.~\ref{fig:schematics}c, the atom-centered representation is made by 
summing over all the edges of a node,
\begin{equation}
    A_{i, cn\mathbf{l}} = 
    \sum_{j\in \mathcal{N}(i)} \chi_{cn\mathbf{l}} (\sigma_i, \sigma_j).
\end{equation}
This corresponds to a projection of the edge basis on the atomic density, which is commonly referred to as the ``density trick'' and was
introduced originally to construct SOAP and bispectrum descriptors~\cite{bartok2013representing}.

\textbf{Symmetrized basis ($B$-basis)}
The orientation-dependent $A$ features are symmetrized to get the invariant $B$ features of different body orders $\nu$ as sketched in Fig.~\ref{fig:schematics}c.
For $l=0$, $A$ is already rotationally invariant, so the $\nu=1$ features $B^{(1)}_{i, cn0} = A_{i, cn\textbf{0}}$.
For $\nu = 2$~\cite{Zhang2021}, 
\begin{equation}
    B_{i, cnl_1}^{(2)} = 
    \sum_{\mathbf{l}_1} 
    \mathcal{C}(\mathbf{l}_1)
    A_{i, cn\mathbf{l}_1}^2.
\end{equation}
Doing the expansion explicitly and using Eqn.~\eqref{eq:l-add}, one can show that~\cite{Zhang2021}
\begin{multline}
        B_{i, cnl_1}^{(2)} = 
   \sum_{j,k\in \mathcal{N}(i)}
    T_{c}(\sigma_i, \sigma_j) R_{n,cl_1}(r_{ji}) 
    T_{c}(\sigma_i, \sigma_k)R_{n,cl_1}(r_{ki}) \\
    \times 
    cos^{l_1}(\theta_{ijk}).
\end{multline}
So this term includes three-body contributions of $\vec{\mathbf{r}}_{ji}$ and $\vec{\mathbf{r}}_{ki}$ together with their enclosed angle $\theta_{ijk}$.
For $\nu = 3$,
\begin{equation}
    B_{i, cnl_1l_2}^{(3)} =
    \sum_{\mathbf{l}_1,\mathbf{l}_2}
\mathcal{C}(\mathbf{l}_1)
\mathcal{C}(\mathbf{l}_2)
    A_{i, cn\mathbf{l}_1} 
    A_{i, cn(\mathbf{l}_1 + \mathbf{l}_2)}
    A_{i, cn\mathbf{l}_2},
\end{equation}
which includes four-body interaction terms in the form
$R_{n,cl_1}(r_{ji})R_{n,c(l_1+l_2)}(r_{ki})R_{n,cl_2}(r_{hi}) 
cos^{l_1}(\theta_{ijk}) cos^{l_2}(\theta_{ikh})$.
For $\nu = 4$,
\begin{multline}
    B_{i, cnl_1l_2l_3}^{(4)} =
    \sum_{\mathbf{l}_1,\mathbf{l}_2,\mathbf{l}_3}
\mathcal{C}(\mathbf{l}_1)
\mathcal{C}(\mathbf{l}_2)
\mathcal{C}(\mathbf{l}_3)\\
\times
 A_{i, cn\mathbf{l}_1} 
 A_{i, cn(\mathbf{l}_1+\mathbf{l}_2)}
A_{i, cn(\mathbf{l}_2+\mathbf{l}_3)} 
A_{i, cn\mathbf{l}_3},
\end{multline}
which includes five-body contributions in the form 
$R_{n,cl_1}(r_{ji})  
R_{n,c(l_1+l_2)}(r_{ki})
R_{n,c(l_2+l_3)}(r_{hi})
R_{n,cl_3}(r_{gi})\\
cos^{l_1}(\theta_{ijk}) 
cos^{l_2}(\theta_{ikh})
cos^{l_3}(\theta_{ihg})$.

\rrev{The general rule for constructing the symmetrized basis is illustrated in Fig.~\ref{fig:schematics}i.
The key idea is that pairs of angular features $L$ need to have a shared factor of $\mathbf{l}$, when summed over to form invariants.
To eliminate linearly dependent terms, the combination of the angular features that only differ from others by permutations are not considered.
Moreover, the shared factor $\mathbf{l}$ between any pairs of $L$ needs to be greater than zero,
because if any $\mathbf{l} = \mathbf{0}$ the resulting invariant can be constructed using the product of lower-order invariants and can thus be eliminated for redundancy.
In other words, a valid combination illustrated in Fig.~\ref{fig:schematics}i needs to correspond to one connected graph, rather than two disconnect graphs of lower body orders.
For instance, terms such as $B_{i, cn012}^{(4)} = B_{i, cn0}^{(1)}B_{i, cn12}^{(3)}$, $B_{i, cn102}^{(4)} = B_{i, cn1}^{(2)}B_{i, cn2}^{(2)}$ are eliminated.
}
The CACE framework thus makes polynomially independent invariants.
Other radial or edge terms that do not depend on $\mathbf{l}$ or only depend on the total angular momentum $l$ can be multiplied on top without breaking the rotational symmetry.

Based on the expansions, one can see that these CACE $B^{(\nu)}$ terms are \rrev{similar to the $B^{(\nu)}$ terms in the original ACE paper~\cite{drautz2019atomic} \rrev{or the invariant basis functions in MTP~\cite{shapeev2016moment}}, 
with two differences:
First, many linearly or polynomially-dependent features are elimiated during the symmetrization stage in CACE, resulting in a much more compact basis set than \rrev{the original ACE}~\cite{drautz2019atomic} or MTPs~\cite{shapeev2016moment}, where the basis functions span the space of all invariants including linearly dependent terms~\cite{dusson2022atomic}.
Second, since each radial channel $R_{n,cl}$ contains all raw radial channels with trainable weights (Eqn.~\eqref{eq:r-mix}), it is not necessary to couple different radial channels together when forming the basis $B$. 
This means that the number of radial channels of $B$ stays $n$ rather than going up as $n^\nu$ as the body order increases.}

\begin{figure}
  \centering
  \includegraphics[width=0.45\textwidth]{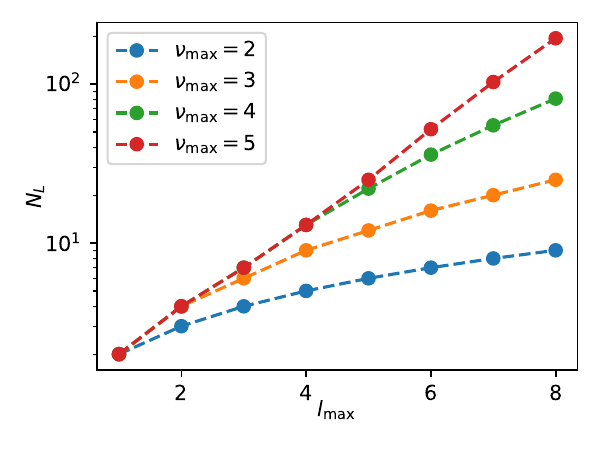}
    \caption{
    The total number of angular features $N_L$, for maximum values of $l_\mathrm{max}$ and $\nu_\mathrm{max}$.
    }
    \label{fig:NL}
\end{figure}

The $B$ basis is a
complete basis of permutation and rotation invariant functions of the
atomic environment~\cite{drautz2019atomic}.
In practice, one truncates the series up to some maximum values of $l_\mathrm{max}$ and $\nu_\mathrm{max}$ to make the computation tractable.
The dimension of the $B$ basis is $c \times n \times N_L$, where each factor comes from the number of edge channels, radial basis, and angular basis, respectively.
Fig.~\ref{fig:NL} shows the total number of angular features $N_L$ up to certain $l_\mathrm{max}$ and $\nu_\mathrm{max}$.
The number of $N_l$ is compact even for  $\nu_\mathrm{max}=5$.
For most practical applications, $l_\mathrm{max}$ between 2 and 4 and $\nu_\mathrm{max}$ between 2 and 4 are sufficient for good accuracy when fitting CACE potentials.

\rev{Notice that in the CACE formulation, each invariant $B$ feature can be evaluated independently,
and there is no need to precompute or store the high-dimensional many-body product basis in
ACE~\cite{drautz2019atomic} or moment tensors in MTPs~\cite{shapeev2016moment} even in the sparse form.
The reason why this is efficient can be understood as
it effectively exploits the extreme sparsity of the Clebsch-Gordan coupling matrix when forming invariants.}

\textbf{Message passing}
The framework above reproduces the original ACE, although formulated entirely in Cartesian space. 
In ACE, the atomic energy of each atom is determined by neighboring atoms within its cutoff radius $r_\mathrm{cut}$.
On top of that, one can incorporate message passing, 
which iteratively propagates information along the edges, 
thereby communicating the local information and effectively enlarging the perceptive field of each atom.
\rev{A systematic discussion on the design choices of message passing can be found in Refs.~\cite{batatia2022design,bochkarev2023atomic}.}

CACE uses two types of message passing (MP) mechanisms.
The first type, denoted by the green arrow in Fig.~\ref{fig:schematics}e,
communicates the $A$ information between neighboring atoms:
the orientation-dependent message from atom $j$ to $i$ can be expressed as:\begin{equation}
    m_{1, j \rightarrow  i, cn\mathbf{l}}^{(t)} = 
    F_{cnl}(r_{ji})
    A_{j, cn\mathbf{l}}^{(t)}
\end{equation}
where $F_{cnl}(r_{ji})$ is a filter function that takes the scalar value of the distance $r_{ji}$, and it can depend on edge channel, radial channel and total angular momentum.
\rev{In practice, we use an exponential decay function with trainable exponent,
to capture the physics that farther atoms have less effect,
although other choices are possible.}

The second MP mechanism, denoted by the two blue arrows in Fig.~\ref{fig:schematics}e,
uses a recursive edge embedding scheme \rev{analogous to recursively embedded atom neural networks (REANN)~\cite{Zhang2021} and multilayer ACE (ml-ACE)~\cite{bochkarev2022multilayer}:}
\begin{equation}
    m_{2, j \rightarrow  i, cn\mathbf{l}}^{(t)} = 
    H_{cl}(B_j^{(t)})
    \chi_{cn\mathbf{l}},
\end{equation}
where $H_{cl}$ is a trainable scalar function that can optionally depend on edge channel and total angular momentum. We simply use a linear neural network (NN) layer for $H$.

The aggregated message is
\begin{equation}
    M_{i, cn\mathbf{l}}^{(t)} =
    \sum_{j\in \mathcal{N}(i)}
m_{1,j \rightarrow  i, cn\mathbf{l}}^{(t)}
+ m_{2,j \rightarrow  i, cn\mathbf{l}}^{(t)}
\end{equation}
One can then update the $A$ representation of each node,
\begin{equation}
    A_{i, cn\mathbf{l}}^{(t+1)}
    = G(A_{i, cn\mathbf{l}}^{(t)}, M_{i, cn\mathbf{l}}^{(t)}),
\end{equation}
and for $G$ we use a simple linear function.
After obtaining the updated $A^{(t+1)}$, we again take the symmetrized $B$-basis as previously described.
It is easy to verify that these symmetrized representations are dependent only on the scalar values of the atomic distances and angles and thus rotationally invariant. 
Note that one can make various design choices for the $F$, $H$ and $G$ functions: they can be dependent on $c$, $n$, and $l$, without breaking the invariance at the symmetrization stage.

In practice, only one or no MP layer is typically used in CACE.
This is because the use of higher-body-order atomic features and messages reduces the need for many MP layers, as in the case of MACE~\cite{batatia2022mace}.

\textbf{Readout and output}
After $T$ MP layers, 
all the resulting  $B$ features at each layer are concatenated.
These invariant features are compact; the number of the features can be computed as $T \times c \times n \times N_L$.
Then, a multilayer perceptron (MLP) maps these invariant features to the target of the atomic energy of each atom $i$,
\begin{equation}
    E_i = MLP(B_i^{(0)} \oplus B_i^{(1)} \oplus \ldots \oplus  B_i^{(T)}).
\end{equation}
In practice, for the readout function, we use the sum of a linear NN and a multilayer NN, as the former preserves the body-ordered contributions and the latter considers the higher-ordered terms from the truncated series.

Finally, the total potential energy of the system is the sum of all the atomic energies of each atom, and the forces can readily be computed by taking the derivatives of the total energy with respect to atomic coordinates.

\begin{table*}
\centering
  \caption{
    Root mean square errors (RMSE) for energy (E) per water molecule (in meV/H$_2$O) and force (in meV\AA$^{-1}$) on a liquid water dataset from Ref.~\cite{cheng2019ab}.
    The cutoffs $r_\mathrm{cut}$ of the atomic environment and the numbers of message passing layers $T$ are listed.
    The NequIP and DeepMD models are trained by us, and the others are from the references.
  }
\begin{tabular}{l c c c c c c c c c} 
  \hline
      & BPNN~\cite{cheng2019ab} 
      & DeepMD~\cite{wang2018deepmd}
      & EANN~\cite{zhang2021accelerating}  
      & linear ACE~\cite{witt2023acepotentials}
      & REANN~\cite{Zhang2021} 
      & NequIP~\cite{batzner20223} 
      & MACE~\cite{batatia2022mace} 
      & CACE
      & CACE
      \\
      & $6.35$~\AA   
      & $6$~\AA   
      & $6.2$~\AA 
      & $5.5$~\AA 
      & $6.2$~\AA, $T=3$ 
      & $4.5$~\AA, $T=3$ 
      & $6$~\AA 
      & $5.5$~\AA, $T=0$
      & $5.5$~\AA, $T=1$
      \\
\hline
E & 7.0  & 6.3 & 6.3 & 5.196 & 2.4  & 2.8 & 1.9  & 3.49 & 1.77 \\ 
F & 120  & 92  & 129 & 99  & 53.2 & 45  & 36.2 & 79  & 47 \\ 
\hline
\end{tabular}
\label{tab:water}
\end{table*}

\subsection{Bulk water}
As an example application on complex fluids,
we demonstrate CACE on a dataset of 1,593 liquid water
configurations~\cite{cheng2019ab}. 
The details of the training are in the Methods section.
The root mean square errors (RMSE) of the energy per water molecule and forces
on the validation set for the CACE model and other MLIPs are presented in Table~\ref{tab:water}.
MLIPs based on three-body and three-body descriptors without message passing, including BPNN based on ACSF~\cite{behler2007generalized}, embedded atom neural network (EANN)~\cite{zhang2021accelerating}, and DeepMD have similar errors, and are amongst the highest in this comparison.
Linear ACE uses higher-body order features,
which can be directly compared to the CACE model trained with the same cutoff and without message passing ($r_\mathrm{cut}=5.5$~\AA, $T=0$),
and the latter has lower error.
The REANN potential~\cite{Zhang2021}
with three message passing layers, the equivariant message passing model NequIP~\cite{batzner20223},
the high body order MACE~\cite{batatia2022mace} model, and the CACE model with $T=1$ have the lowest errors.

\begin{figure}[h]
  \centering
 \includegraphics[width=0.45\textwidth]{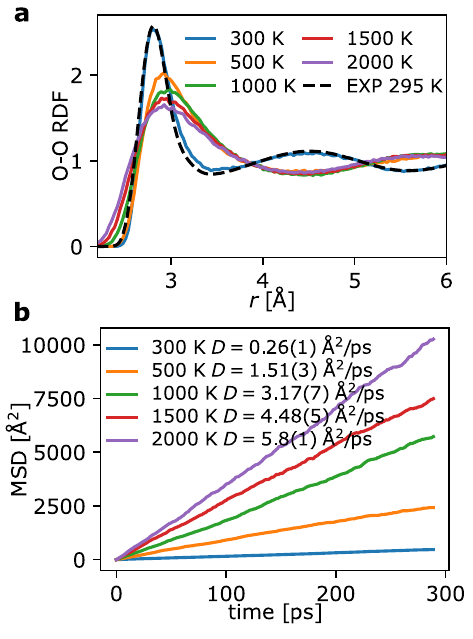}
    \caption{
    Simulation results of water using CACE $T=1$ model.
    \figLabelCapt{a} 
    Oxygen-oxygen radial distribution functions (RDF) at different temperatures and 1~g/mL computed via classical MD simulations in the NVT ensemble.
    The experimental O-O RDF at ambient conditions was obtained from Ref~\cite{skinner2014structure}.
    \figLabelCapt{b} Mean squared displacement (MSD) from  the liquid water simulations.
    Diffusivities ($D$) are shown in the legends.
    }
    \label{fig:water}
\end{figure}

For running molecular dynamics (MD) simulations using MLIPs,
besides accuracy, a crucial requirement is stability,
the possibility to run long enough trajectories with the system remaining in physically reasonable states and the simulation not exploding, even at conditions with few or no training data.
To assess the simulation stability, we performed independent simulations of 300~ps in length for water at 1~gmL$^{-1}$ at 300~K, 500~K, 1000~K, 1500~K, and 2000~K using the CACE $T=1$ model. 
The simulation cell contains 512 water molecules.
The time step was 1 femtosecond and the Nosé-Hoover thermostat was used.
The upper panel of Fig.~\ref{fig:water} shows the oxygen-oxygen (O-O) radial distribution function (RDF).
At 300~K, the computed O-O RDF is in excellent agreement with experiment from X-ray diffraction
measurements~\cite{skinner2014structure}.
Note that the nuclear quantum effects have a slight destructuring effect in the O-O RDF, but the effect is quite small~\cite{cheng2019ab}.
The mean squared displacements (MSD) from these simulations
are shown on the lower panel of Fig.~\ref{fig:water}.
The diffusivity $D$, obtained by fitting a linear function to the MSD,
at 300~K agrees well with the DFT MD result ($D_\mathrm{DFT} = 2.67 \pm 0.10 \mathrm{\AA}^2$ps$^{-1}$) from Ref.~\cite{marsalek2017quantum}.
At high temperatures up to 2000~K, the MD simulations remained stable, which demonstrates the superior stability of the CACE potential.
Note that we do not expect the result at very high temperatures to be physically predictive, and this example is just to illustrate the stability of CACE.
Such stability is not only necessary for converging statistical properties,
but also allows the refinement of the potential: 
to iteratively improve the accuracy of the potential under certain conditions, one can perform active learning which collects snapshots from corresponding MD trajectories and use these snapshots for re-training.

\subsection{Small molecules: ethanol and 3BPA}

To demonstrate CACE on small organic molecules, we fitted to 1,000 ethanol structures from the MD17 database~\cite{chmiela2017machine} that is often used for benchmark purposes.
Table~\ref{tab:ethanol} shows a comparison of the different MLIPs trained on energies and forces for the same MD17-ethanol dataset.
The energy accuracy of CACE is similar to the NequIP model with  $T=5$, although the CACE force error is larger.
The stability (S) in Table~\ref{tab:ethanol} measures how long MD simulations at 500~K using the MLIP is physically sensical (no pathological behavior or enter physically prohibitive states) in runs with a maximum length of 300~ps, so 300~ps is the maximum score in this case~\cite{fu2023forces}.
The stability values for other MLIPs are from Ref.~\cite{fu2023forces} which used 10,000 training structures and a time step of 0.5~fs in MD.
For CACE, we used 1,000 training structures and a time step of 1~fs, which is more stringent, but the MD had nevertheless remained stable.
As observed in Refs.~\cite{fu2023forces,stocker2022robust}, force accuracy does not imply stability, and the latter is a key metric for MLIPs.

\begin{table*}
\centering
\begin{tabular}{lccccccccc}
\hline 
     & SchNet~\cite{schutt2017schnet} & DimeNet~\cite{gasteiger2019directional} & sGDML \cite{chmiela2019sgdml}& PaiNN~\cite{schutt2021equivariant} 
     & SphereNet~\cite{liu2022spherical} & NewtonNet~\cite{haghighatlari2022newtonnet}  & GemNet-T~\cite{gasteiger2021gemnet}  & NequIP~\cite{batzner20223} 
     & CACE\\
\hline 
E    & 3.5  & 2.8  & 3.0  & 2.7 &     & 2.6 &     & 2.2 & 2.37 \\  
F    & 16.9 & 10.0 & 14.3 & 9.7 & 9.0 & 9.1 & 3.7 & 3.1 & 7.0 \\    
S $\uparrow$    & 247  &  26  &      & 86  & 33  &     & 169 & 300 & 300  \\
\hline 
    \end{tabular}
    \caption{Mean absolute error (MAE) of energy (E) and force (F) for ethanol in the MD-17 dataset, in units of meV and meV\AA$^{-1}$,
respectively, trained on 1000 reference configurations.
The stability (S) in the unit of ps 
shows how long the simulation using the MLIP is physically realistic in MD of a maximum length of 300 ps at 500~K.
Besides CACE, the errors are obtained from respective references, and the stability is from Ref.~\cite{fu2023forces}.}
    \label{tab:ethanol}
\end{table*}

\begin{table*}
\centering
\begin{tabular}{lcccccccc}
\hline 
            & ACE  \cite{drautz2019atomic} & sGDML \cite{chmiela2019sgdml}    & GAP \cite{bartok2010gaussian}         & ANI-pretrained \cite{gao2020torchani}   & ANI-2x \cite{gao2020torchani}    & NequIP~\cite{musaelian2023learning}            & MACE~\cite{batatia2022mace} & CACE \\
\hline 
300 K, E    & 7.1  & 9.1        & 22.8  & 23.5      &  38.6     &  3.3 & 3.0  &  6.3 \\  
300 K, F    & 27.1 & 46.2       & 87.3  & 42.8      & 84.4      &  10.8 & 8.8  & 21.4 \\     
\hline
600 K, E    & 24.0 & 484.8      & 61.4  & 37.8      &  54.5     & 11.2 & 9.7  & 18.0 \\       
600 K, F    & 64.3 & 439.2      & 151.9 & 71.7      & 102.8     &  26.4 & 21.8 &  45.2 \\       
\hline 
1200 K, E   & 85.3 & 774.5      & 166.8 & 76.8      & 88.8      &  38.5 & 29.8 & 58.0\\             
1200 K, F   & 187.0  & 711.1    & 305.5 & 129.6     & 139.6     &  76.2 & 62.0 & 113.8 \\ 
\hline 
    \end{tabular}
    \caption{Root mean square errors (RMSE) for energy (E) and force (F) on the 3BPA temperature transferability dataset, reported in units of meV and meV\AA$^{-1}$. 
    All models were trained on $T=300$~K data. Results of previous models except for NequIP and MACE are from \cite{kovacs2021linear}.
    ANI-2x was trained on 8.9 million structures,
    and  ANI-pretrained was pretrained and fine-tuned on this dataset.}
    \label{tab:3bpa}
\end{table*}

\begin{figure}[h]
  \centering
 \includegraphics[width=0.45\textwidth]{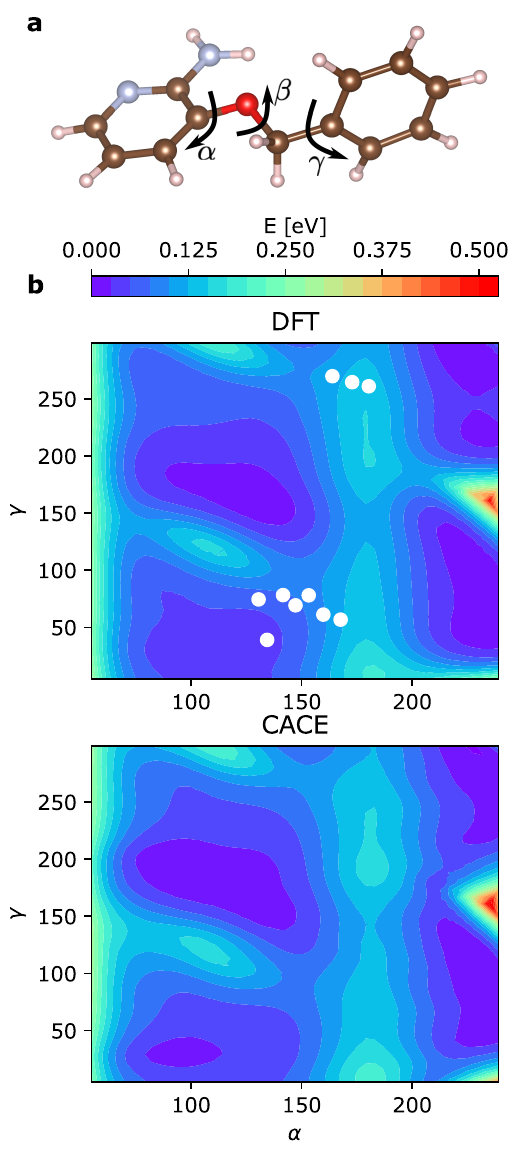}
    \caption{
    The dihedral scan benchmark of the 3BPA molecule.
    \figLabelCapt{a} Molecular structure and three dihedral angles.
    \figLabelCapt{b} 
    The dihedral potential energy landscape for $\beta=120^{\circ}$, as predicted by DFT and CACE. The white dots on the DFT potential energy surface correspond to all the training configurations that have $\beta$ between 100$^{\circ}$ and 140$^{\circ}$.
    }
    \label{fig:bpa}
\end{figure}

As another example,
we fitted to the 3BPA (3-(benzyloxy)pyridin-2-amine) dataset~\cite{kovacs2021linear}.
Table~\ref{tab:3bpa} shows the errors of different potentials when extrapolating to higher temperature.
NequIP 
and MACE 
perform the best,
while CACE is somewhere between them and ACE.
We then compared the dihedral potential
energy surface of the molecule.
Fig.~\ref{fig:bpa} shows the complex energy landscapes of the $\alpha$-$\gamma$ plane predicted by DFT and by CACE, for the case of $\beta=120^{\circ}$, the plane with the fewest training points.
Nevertheless,
the CACE landscape closely resembles the ground truth and is regular, more so than the other potentials benchmarked in Ref.~\cite{kovacs2021linear} despite those were trained on a more diverse dataset including high temperature configurations.

\subsection{25-element high-entropy alloys}

High-entropy alloys, which are composed of several metallic elements, 
have unique mechanical, thermodynamics, and catalytic properties~\cite{george2019high}.
They are challenging to model due to the high number of elements involved, for which many MLIPs scale poorly with.
Ref.~\cite{lopanitsyna2023modeling} introduced a 25 d-block transition metal HEA dataset of distorted crystalline structures.
Ref.~\cite{lopanitsyna2023modeling} further performed an alchemical learning (AL) that explicit compresses chemical space, leading to a model that is both interpretable and very stable.

\begin{table}
\centering
\begin{tabular}{lccc}
\hline 
 & AL 3B+NN~\cite{lopanitsyna2023modeling}   & PET~\cite{pozdnyakov2023smooth} & CACE\\
\hline 
test, E   & 10  & 1.87 & 5.28 \\  
test, F   & 190 & 60.1 & 111 \\ 
\hline
substitute, E & 24  & N.A. & 8.5 \\  
substitute, F & 373 & N.A. & 124 \\ 
\hline
5000~K, E & 48  & 152 & 15.3 \\  
5000~K, F & 290 &     & 233 \\ 
\hline
    \end{tabular}
    \caption{Mean absolute error (MAE) of energy (E) and force (F) for the HEA25
    dataset, reported in units of meV and meV\AA$^{-1}$. 
    There are three sets of benchmarks on test error, 
    alchemical extrapolation to unseen Re and Os substitute atoms,
    and temperature transferability to 5000~K.}
    \label{tab:hea}
\end{table}

\begin{figure*}[h]
  \centering
  \includegraphics[width=0.8\textwidth]{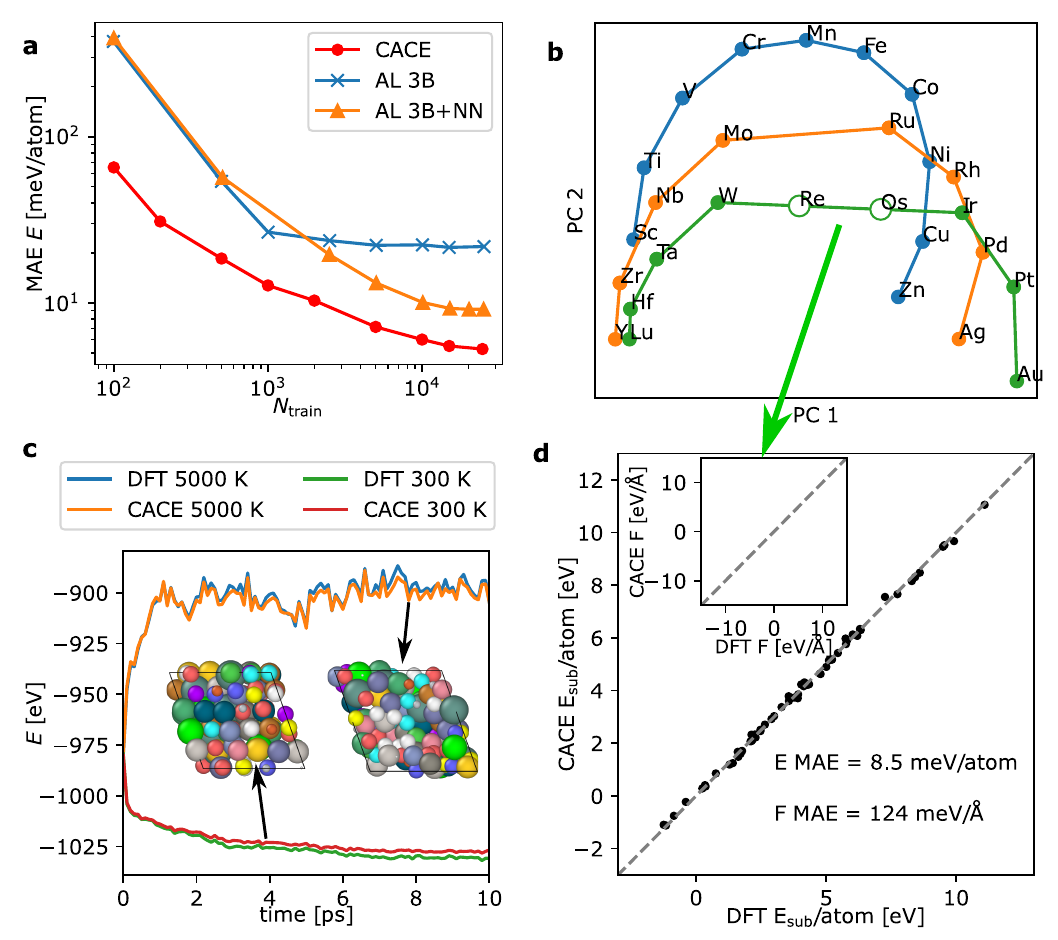}
    \caption{
    Benchmark on the HEA25 dataset.
    \figLabelCapt{a} 
    Learning curves for different models.
    Both alchemical learning (AL) models are from Ref.~\cite{lopanitsyna2023modeling}.
    \figLabelCapt{b}
    The first two principal components (PCs) of the CACE embedding matrix. 
    The periods are highlighted with orange, blue, and green lines. Interpolated positions for Re and Os are indicated with empty circles. 
    \figLabelCapt{c}
    Energy comparison between DFT and CACE for a high temperature (5000~K) and a low temperature (300~K) trajectory.
    The inset shows example solid and melt configurations from the 300~K and the 5000~K trajectories, respectively.
    \figLabelCapt{d}
    Comparison between the substitution energy $E_\mathrm{sub}$, the potential energy difference between the original structure and the structure substituted with Re and Os element per substitution atom, computed using DFT and CACE. The inset shows the parity plot for the force
components computed for those structures. 
    }
    \label{fig:hea}
\end{figure*}

In Fig.~\ref{fig:hea}a, the CACE learning curve is compared to
the ones from Ref.~\cite{lopanitsyna2023modeling},
including a 3B model that has an atomic energy baseline and three-body terms,
and another 3B+NN model that further incorporates a full set of pair potentials and a non-linear term built on top of contracted power spectrum features.
Overall, the CACE model achieved lower error and significantly more efficient learning for fewer data,
and it has only
91,232
trainable parameters, compared to  more than 160,000 in the 3B+NN model~\cite{lopanitsyna2023modeling}.

The errors of the models are presented in Table~\ref{tab:hea}.
A very recent architecture, Point Edge Transformer (PET) that utilizes local coordinate systems, achieved state-of-the-art performance in several benchmark systems and also has impressively low test error for the HEA25 dataset~\cite{pozdnyakov2023smooth}.
However, as we discuss below, the PET potential has poor transferability.

Alchemical learning capability not only makes learning more efficient, but is also essential towards achieving a foundation model that is valid across the periodic table.
CACE is able to perform alchemical learning. 
The learnable embedding $\boldsymbol{\theta}$ for each element type encodes its chemical information,
and can be visualized to gain insights into the data-driven similarities.
As the $N_\mathrm{embedding}=4$ in this case, we performed a principal component analysis (PCA) and plotted the first two principal component axes in Fig.~\ref{fig:hea}b.
The elements are arranged in a way that is strongly
reminiscent of their placement in the d-block.
This shows that the element embedding scheme is able to learn the nature of the periodic table in a data-driven way.

A stringent way to demonstrate alchemical learning is by extrapolating to unseen elements.
In this case, Re and Os are absent from the training set.
Ref.~\cite{lopanitsyna2023modeling} provides a hold-out set containing 61 pairs of structures, each of 36 or 48 atoms. 
Each pair has one fcc or bcc structure with only the original 25 elements,
and another structure with up to 11 atoms randomly substituted with Re and Os.
We simply took the CACE model trained for the 25-element dataset
and obtained the chemical embedding of Re and Os by a linear interpolation between W and Ir,
as illustrated by the empty circles in Fig.~\ref{fig:hea}b.
We also added atomic-energy baselines
for Re and Os (obtained by fitting the residual energy to
a two-parameter linear model that depends only on
the Re and Os content).
In Fig.~\ref{fig:hea}d we show the parity plot for the substitution energy, defined as the potential energy difference between the original and the substituted structure per substitution atom.
Impressively, the CACE error for those substituted structures with unseen elements is rather similar to the test error on the 25-element training set,
and is lower than the AL method in Ref.~\cite{lopanitsyna2023modeling}.
The force comparison, shown in the inset of Fig.~\ref{fig:hea}d,
shows the same trend.
With the PET potential, it is not clear whether alchemical learning can be performed in a straightforward way.

The most impressive validation of the AL model in Ref.~\cite{lopanitsyna2023modeling} is the extrapolation to configurations very far from the training set:
At 5000~K, the systems generated from constant-pressure MD simulations with Monte Carlo element swaps have melted,
but the AL potential that was trained exclusively on
solid structures is still able to describe the energetics of the structures.
This is a test that the PET model~\cite{pozdnyakov2023smooth} did not perform well.
To verify the generalizability of CACE across a wide temperature range,
we recomputed the energies and forces of the structures from the 5000~K trajectory as well as another one at a low temperature of 300~K,
and show the results in Fig.~\ref{fig:hea}c and Table~\ref{tab:hea}. 
The CACE errors are a bit smaller compared to Ref.~\cite{lopanitsyna2023modeling},
and quite modest compared to the training error.

\section{Discussion}

There have been other MLIP approaches that only operate in Cartesian space, 
including NewtonNet~\cite{haghighatlari2022newtonnet}, EGNN~\cite{satorras2021n}, and PaiNN~\cite{schutt2021equivariant} that use atomic displacement vectors in message passing,
the PET based on local coordinate systems~\cite{pozdnyakov2023smooth},
and TensorNet that uses Cartesian tensors to represent pairwise displacements~\cite{simeon2023tensornet}.
CACE is different from these approaches in that it can be directly mapped to ACE, so like ACE, it is also complete and body-ordered.
The completeness helps the accuracy and the smoothness of the potential,
particularly when there exist atomic environments that are degenerate using three-body descriptors or bispectrum~\cite{pozdnyakov2020incompleteness,nigam2023completeness}.
The body order makes the potential more physically interpretable and may help extrapolation.
\rev{The angular basis of CACE (Eqn.~\eqref{eq:angular-basis}) is the same as MTP~\cite{shapeev2016moment}, but CACE avoids computing moment tensors, and form a more compact symetric basis set via a simple combination rule that eliminates polynomially depedent features (Fig.~\ref{fig:schematics}i).}
\rrev{As added advantages, the CACE descriptors have optimized radial channel coupling, and are capable of alchemical learning.}

A topic that tends to be overlooked in the development of MLIP methods is the stability and generalizability.
In practice, a MLIP that is not stable in MD simulations is not useful.
Nonetheless, stability and generalization are rarely considered in benchmarks, maybe due to that they are less straightforward to assess than energy and force errors.
On the other hand, it has been shown that accuracy of the potential does not imply stability~\cite{fu2023forces,stocker2022robust}. 
The CACE potential shows high stability and extrapolatability in the datasets we benchmarked:
stable MD simulations of water up to 2000~K,
MD of ethanol, extrapolation to high temperature and dihedral scan of 3BPA,
as well as generalization to the melt and unseen elements in high-entropy alloys.
For water, the accuracy of CACE is on par with the most accurate potentials to date, NequIP~\cite{batzner20223} and MACE~\cite{batatia2022mace}.
For small molecules, CACE is less accurate than NequIP or MACE but more accurate than the other recent MLIPs that we compared to.
For all practical purposes, the accuracy of CACE for the small molecules is probably enough, considering that the error is very small in absolute terms and is much lower than the chemical accuracy.
Moreover, CACE achieved such accuracy with only one message passing layer and relatively few training parameters, and its accuracy can probably be improved with more message passing layers, higher $N_\mathrm{embedding}$, $\nu_\mathrm{max}$, $l_\mathrm{max}$,
or a larger MLP at the readout.

\begin{table}
\centering
\begin{tabular}{lccc}
\hline 
 & NequiIP~\cite{batzner20223}   & MACE~\cite{batatia2022mace} & CACE\\
       & $4.5$~\AA, $T=3$ 
      & $6$~\AA 
      & $5.5$~\AA, $T=1$ \\
\hline 
1,536 atoms & 20 ps per h & 18 ps per h & 24 ps per h \\
5,184 atoms & 7 ps per h & 4.8 ps per h & 8 ps per h\\
12,288 atoms & 3 ps per h & 2 ps per h & 3.5 ps per h\\
24,000 atoms & 1.65 ps per h & N.A. & 1.7 ps per h \\
\hline
    \end{tabular}
    \caption{Speed of MD of liquid water (in picosecond of simulation per hour) using different water models on a single Nvidia A100 GPU.}
    \label{tab:speed}
\end{table}

We implemented the CACE potential using PyTorch, and the code is available in 
    \url{https://github.com/BingqingCheng/cace}.
\rev{With the current implementation, the training time of the water and the HEA25 datasets took about two days on a vintage GeForce GTX 1080 Ti GPU card,
and the small molecules can be trained on a laptop within one or two days.
The training should be much faster on state-of-the-art GPUs.
For the speed of MD,
Table~\ref{tab:speed} compares the timing on one Nvidia A100 GPU using three water models in Table~\ref{tab:water}.
Note that such comparison depends very much on the specific models used (higher $\nu$, $l_\mathrm{max}$, $r_\mathrm{cut}$, and $T$ all add to the cost) as well as the implementation.
CACE is marginally faster in this comparison and can drive simulations of 24,000 atoms.
The ability to run such large system sizes on a single GPU speaks to the low memory consumption of CACE.}
The implementation of the code can be further optimized:
The computation of the node features and the symmetrization step may be compiled and be made faster.
The training and the prediction are available on one GPU and can be made more parallelizable.
The MD simulations can be performed in the ASE package, but the LAMMPS interface implementation is currently absent.
As CACE has a moderate receptive field,
so it is in principle possible to make it scale well in MD simulations.

Many engineering aspects of the CACE potential can be further fine-tuned. For example, dataset normalization and internal normalization can have a large effect on training efficiency and outcome~\cite{batatia2022design}.
Radial basis influences the learning efficiency~\cite{bigi2022smooth},
and its best choice is still an open question.
Other choices of the
$F$, $H$ and $G$ 
functions in the message passing may be used.
One can also use the product of $B_i^{(t)})$
and $\chi_{cn\mathbf{l}}$ in the message function, exploiting the relationship in Eqn.~\eqref{eq:l-add}.
The training protocol can also influence the outcome.

Other possible future improvements of CACE include pretraining on large datasets before fine-tuning the model. It should be possible to develop a foundation model that is generally valid in the periodic table~\cite{batatia2023foundation}. The alchemical learning capacity of CACE may facilitate this.


In summary, we propose a scheme to preserve the rotational symmetry of atomic environments with different body orders, and the operation is entirely in the Cartesian coordinates.
\rev{The transformation to the spherical harmonic basis, tensor product of atomic basis, and the usage of high-dimensional Clebsch-Gordan coupling matrix are all avoided.
The resulting symmeric features are low-dimensional and polynormially indepedent.}
As the symmetrization in the Cartesian coordinates is equivalent to the Clebsch-Gordan contraction,
the scheme can be applied in other MLIPs that use the ACE framework (e.g. linear ACE~\cite{witt2023acepotentials}, SNAP, SOAP-GAP~\cite{bartok2010gaussian}, Jacobi-Legendre cluster expansion~\cite{domina2023cluster}, ml-ACE~\cite{bochkarev2022multilayer}, \rev{graph ACE~\cite{bochkarev2023atomic})}, or E(3) equivariant message passing neural networks (e.g. NequIP~\cite{batzner20223}, MACE~\cite{batatia2022mace}, 
BOTNet~\cite{batatia2022design}, SEGNN~\cite{brandstetter2021geometric}, Allegro~\cite{musaelian2023learning}).

\section{Methods}

\textbf{Bulk water}
The water dataset has 1,593 liquid water
configurations, each of 64 molecules~\cite{cheng2019ab}. 
It was computed at
the revPBE0-D3 level of density functional theory (DFT).
90\% of the data are used in the training.
We used a cutoff $r_\mathrm{cut}=5.5$ \AA, 6 Bessel radial functions and $c=12$, $l_\mathrm{max}=3$, $\nu_\mathrm{max}=3$, $N_\mathrm{embedding}=3$, and no message passing ($T=0$) or one message passing layer ($T=1$).
The fitted CACE models have relatively few parameters: 24,572 trainable parameters for the CACE water model with $T=0$, and
69,320
trainable parameters for the model with $T=1$.

In Table~\ref{tab:water}, the NequIP and DeepMD models are trained by us, and the others are from the references.
In addition, we also trained a MACE model with E RMSE=1.9~meV/H$_2$O, F RMSE=46.8~meV\AA$^{-1}$.
as the original Behler-Parrinello neural network (BPNN) in Ref.~\cite{cheng2019ab} was trained using the RuNNer code~\cite{runner}, we also trained another one using N2P2~\cite{singraber2019parallel} and obtained E RMSE=5.8~meV/H$_2$O, F RMSE=108~meV\AA$^{-1}$.

\textbf{Small molecules}
The ethanol data from the MD17 database~\cite{chmiela2017machine} were sampled from DFT molecular dynamics simulations at 500~K.
We used a cutoff $r_\mathrm{cut}=4$~\AA, 6 Bessel radial functions,
$c=12$, $l_\mathrm{max}=4$, $\nu_\mathrm{max}=3$, $N_\mathrm{embedding}=4$, and one message passing layer ($T=1$).
The model was trained on 1,000 structures randomly selected from the dataset (90\% train/10\% validation split), and tested on another 10,000 random structures.

The 3BPA dataset~\cite{kovacs2021linear} set consists of 500 training structures at 300~K. Test data are from simulations at 300~K, 600~K, and 1200~K.
For fitting the CACE model, we used $r_\mathrm{cut}=6$ \AA, 8 Bessel radial functions,
$c=12$, $l_\mathrm{max}=4$, $\nu_\mathrm{max}=3$, $N_\mathrm{embedding}=4$, and $T=1$.

\textbf{25-element high-entropy alloys}
The HEA dataset from Ref.~\cite{lopanitsyna2023modeling} consists of 25,630 distorted crystalline structures containing 36 or 48 atoms on bcc or fcc lattices.
We took a test set containing 1,000 configurations, and used up to the rest of the  24,630 data points for training and validation.
We used a CACE model with $r_\mathrm{cut}=4.5$~\AA, $l_\mathrm{max}=3$, $\nu_\mathrm{max}=3$, $N_\mathrm{embedding} = 3$, and $T=1$.

\textbf{Data availability}
The water dataset is from
\url{https://github.com/BingqingCheng/ab-initio-thermodynamics-of-water}.
MD17-ethanol is from \url{http://www.sgdml.org/#datasets}.
BPA is from Ref.~\cite{kovacs2021linear}, downloaded from \url{https://github.com/davkovacs/BOTNet-datasets}.
The HEA25 dataset is from Ref.~\cite{lopanitsyna2023modeling}, downloaded from ~\url{https://archive.materialscloud.org/record/2023.57}.

The training scripts, trained CACE potentials, and MD input files are available at \url{https://github.com/BingqingCheng/cacefit}

\textbf{Code availability}
The CACE package is publicly available at \url{https://github.com/BingqingCheng/cace}.

\textbf{Competing Interests}
The author declares no competing financial or non-financial interests.

\textbf{Author Contributions}
BC designed and performed the study, and wrote the paper.

\textbf{Acknowledgements}

BC thanks Ralf Drautz and Ngoc Cuong Nguyen for illuminating discussions.

%

\end{document}